\documentclass[aps,prd,groupedaddress,showpacs,nofootinbib]{revtex4}
\usepackage{graphicx,amssymb,amsmath,epsfig}
\usepackage[english]{babel}
\def\comment#1{}
\newcommand{\be}{\begin{equation}}\newcommand{\ee}{\end{equation}}
\newcommand{\bea}{\begin{eqnarray}}\newcommand{\eea}{\end{eqnarray}}
\newcommand{\beaa}{\begin{eqnarray}}\newcommand{\eeaa}{\end{eqnarray}}
\newcommand{\ba}{\begin{array}}\newcommand{\ea}{\end{array}}
\newcommand{\bit}{\begin{itemize}}\newcommand{\eit}{\end{itemize}}
\newcommand{\ben}{\begin{enumerate}}\newcommand{\een}{\end{enumerate}}

 \newcommand{\sfrac}[2]{\raisebox{0.095ex}{\scriptsize${\frac{#1}{#2}}$}}
 \newcommand{\ssbf}[1]{\mbox{\tiny\bf{#1}}}
 \newcommand{\sbf}[1]{\mbox{\scriptsize\bf{#1}}}

\def\lfrac#1#2{#1/#2}

\begin{document}

%\bibliography{gns}
%\bibliographystyle{plain}

\title{Neutrino Mass Differences
and Nonunitarity of Neutrino Mixing Matrix
from Interfering Recoil Ions
}

\author{H. Kleinert${}^{ab}$ and P. Kienle${}^{cd}$}

%\vspace{2mm}

       \address{
${}^{a}$Institut f\"ur Theoretische Physik,
 Freie Universit\"at Berlin,
Arnimallee 14, D14195 Berlin, Germany\\
$^{(b)}$ICRANeT, Piazzale della Republica 1, 10 -65122, Pescara, Italy\\
    $^c$Stefan Meyer Institut f\"ur
    subatomare Physik, \"Osterreichische Akademie der Wissenschaften,
    Boltzmanngasse 3, A-1090, Wien, Austria\\
   $^d$Physik-Department, Technische Universit\"at M\"unchen,
    D--85748 Garching, Germany}

\vspace{2mm}

\begin{abstract}
We show that the recent observation
of the time modulation of two-body weak decays
of heavy ions
reveals
the  mass content  of the electron
neutrinos
%, without
%detecting
%them directly,
via
interference patterns in the
recoiling ion wave function.
From the modulation period
we derive
the
difference of the square masses
$ \Delta m^2\approx 22.5\times 10^{-5}$\,eV${}^2$,
which is about 2.8 times larger
than that derived from a combined analysis of
KamLAND and solar neutrino oscillation experiments.
It is, however, compatible with a
data regime
to which the KamLAND analysis
attributes a smaller
probability.

The experimental results
imply that the neutrino
mixing matrix violates unitarity by about 10\%.

\end{abstract}

\maketitle
\section{Introduction}

At the GSI in Darmstadt, the experimental
storage ring ESR permits observing
completely ionized heavy atoms $I$  or hydrogen-like
heavy ions $I_H$ over a long time \cite{GSI2,OLD}
and thus to measure the time dependence of
their weak two-body decays
$I_H\rightarrow I+ \nu_e$
or
$I\rightarrow I_H+\bar  \nu_e$.
The first is the
 well-known electron-capture
(EC) process.
The virtue of
such experiments is that
the properties
of the
neutrino or antineutrino
can be deduced
from measurements of
the time dependence of the transition
observing only
the
initial and final ions.
The special efficiency of
these experiments
becomes clear
in the
Dirac
sea interpretation
of the second process, where the initial ion
simply absorbs a negative-energy
antineutrino
in the vacuum.
Since the vacuum  has {\em all\/} negative-energy states filled, the
 vacuum
 is a  source of negative-energy neutrinos of
{\em maximally possible\/}
 current density, i.e.,
the best possible neutrino source in the universe.
This is why
the ESR experiments
yield information on
neutrino properties with great precision
even if the
targets and exposure times are
quite small,
in particular much smaller
than the
$2.44\times 10^{32}$ proton-yrs (2881 ton-yrs)
in the famous KamLAND experiments \cite{Kam},
which are only sensitive
to the much less abundant positive-energy neutrinos
produced by nuclear reactors.

Apart from the neutrino mass difference,
the experiment reveals also another important property of the
presently popular neutrino mixing scheme: the matrix
which expresses the neutrino flavor states into
 fixed-mass states must be  {\em nonunitary\/} to explain the data.
The measurement determines the degree of nonunitarity
to be roughly 10\%.

\section{Two-Neutrino Mixing}

To illustrate this we consider here
at first only
the two lightest
neutrinos.
According to Pontecorvo
\cite{Pontecorvo:1957cp,Wolfenstein:1977ue},
 the Dirac fields
of the physical electron and muon-neutrinos
$ \nu_f =(\nu _e,\nu _\mu)$,
the so-called {\em flavor fields\/},
are superpositions of neutrino fields
  ${\nu}_i =(\nu _1,\nu _2)$
of masses $m_1$ and $m_2$:
\begin{eqnarray} \label{mix1}
\nu_{e}(x) =\,\nu_{1}(x)\,\cos\theta + \nu_{2}(x)\,\sin\theta,~~~~
\label{mix2} \nu_{\mu}(x) &=&\!\!\!-\nu_{1}(x)\,\sin\theta +
\nu_{2}(x)\,\cos\theta \;,\end{eqnarray}
where $\theta$ is a mixing angle.
This is, of course,
the neutrino analog of the
famous Cabibbo mixing of up and down quarks.
The free
Dirac action
has the form
$$
{\cal A}=\sum_f\int d^4x\,\bar \nu_f(x)\left(i \gamma ^\mu \partial _\mu-
{\cal M}\right) \nu_f(x) ,
$$
where
$ \gamma ^\mu$ are the Dirac matrices,
and ${\cal M}$ is a mass matrix, whose diagonal
and
off-diagonal
elements are $m_f=
(m_e,m _\mu)$ and $
m_{e\mu}=
m_{\mu e}$, respectively.
The
eigenvalues
$m_i=(m_1,m_2)$ are related to $m_f$ by
 \cite{Pontecorvo:1957cp,Wolfenstein:1977ue,BV95,BLALO},
\begin{eqnarray}
m_e= m_1\cos^2\theta+ m_2\sin^2\theta,~~~
m_\mu= m_1\sin^2\theta+ m_2\cos^2\theta,~~~\nonumber \\
m_{e\mu}=
m_{\mu e}=
(m_2-m_1)\sin \theta\cos \theta. ~~~~~~~~ ~~~~~~~~~
\label{@}\end{eqnarray}

The weak transition between the electron
$e$ and its neutrino $ \nu _e$
is governed by the interaction
\begin{eqnarray}
{\cal A}_{\rm int}=
{g}\int d^4x\, W^-_\mu(x)J^+{}^\mu(x)+{\rm h}.{\rm c}.
\equiv{g}\int d^4x\, W^-_\mu(x)\,\bar e(x) \gamma ^\mu(1- \gamma _5) \nu _e(x)
+
{\rm h}.{\rm c.},
\label{@WINT}\end{eqnarray}
where $ \gamma _5$ is the product of Dirac matrices $i
 \gamma ^0
 \gamma ^1
 \gamma ^2
 \gamma ^3$.

Since the interaction (\ref{@WINT})
involve only
the flavor
 fields
(\ref{mix1}),
the states  of masses $m_i$
will always be produced as coherent superpositions.
The weakness
of the interaction
will allow us to calculate
the shape of the
mixed
wave packet
from perturbation theory.
Consider the decay
$I\rightarrow I_H+\bar  \nu_e $
which is a superposition of
the states
 of masses $m_1$ and $m_2$.
The formulas will be applicable
for electron capture
if we
exchange $M_H$ by
the mass  $M$ of the bare ion
and deal with outgoing neutrinos.

In the center-of-mass (CM) frame of the initial bare ion
of mass $M$,
the final $H$-like ion
has the same momentum as the antineutrino
$\bar \nu _i$ ($i=1,2$),
whose
energy is
$ \omega _{i}\equiv
\omega _{{\bf k}_i,i}= \sqrt{{\bf k}^2_i+m_i^2} $
determined by
\begin{equation}
M\equiv M_H+Q= \omega _{i} + \sqrt{M_H^2+ {\bf k}_i^2}= \omega _{i}
 + \sqrt{M_H^2+ \omega _{i}^2-m_i^2},~~~i=1,2,
\label{@QEQ}\end{equation}
so that
\comment{
Note that $\bar \nu _1$ and
$\bar \nu _2$
emerge with {\em different\/} momenta
  ${\bf k}_1$ and
  ${\bf k}_2$.
In the experiments
under consideration, $Q$ is
 of the order of
a few MeV, and
the masses $m_i$ are much smaller than $Q$.
}%
\begin{eqnarray}
\omega _{i}
= \lfrac{[(2M_H+Q)Q+m_i^2]}{2(M_H+Q)}.
\label{@OmEn}\end{eqnarray}
Subtracting $ \omega _2$ and $ \omega _1$
from each other
we find the energy difference
\begin{eqnarray}
 \Delta  \omega \equiv   \omega _{2}
  -  \omega   _{1} =
\frac{  m_2^2-m_1^2}{2M }\equiv
 \frac{ \Delta  m^2}{2M }.
\label{@EQUA}
\end{eqnarray}
The denominator $M$ is of the order of 100 GeV
and much larger than $ \Delta  m^2$, so that
$ \Delta  \omega$
is
extremely small.
It is  the difference
of the recoil energies
transferred to the outcoming ion
by the antineutrinos of
masses $m_{1}$
and $m_2$.
Without recoil,
we would have found the
four orders of magnitude
larger
energy difference at the {\em same\/} momentum
%
%\begin{equation}
$  \Delta  \omega_{{\bf k}}=
\omega _{{\bf k},2}
-\omega _{{\bf k},1}
=  (\Delta m^2+ \omega _{{\bf k},1}^2)^{1/2}-
 \omega _{{\bf k},1}
\approx
 \Delta m^2/2 \omega _{{\bf k},1}
\approx  \Delta m^2/2Q.
$ %\label{@ENDI}\end{equation}
This is the frequency with which the incoming
negative-energy
neutrino current
of momentum ${\bf k}$ oscillates
in the vacuum.
Note that although $ \Delta  \omega $
is small, the momentum
difference
$ \Delta k\equiv k_2-k_1$
associated with the energies $ \omega _{1,2}$
is as large as
$  \Delta  \omega_{{\bf k}}$,  but has the opposite sign.
\comment{
We expect the
wave function
of the final state
$I_H+\bar  \nu _e$
to
contain
the
two different momenta
${\bf k}_1$ and
${\bf k}_2$
 associated with the
frequencies
$ \omega _1 $ and $ \omega _2$.
}
\comment{Indeed, if we use the fact that
in the experiment
 $ m^2_i\ll |{\bf k}|^2$, we can approximate
$ \omega _{{\bf k},i}\approx |{\bf k}_i|+m_i^2/2|{\bf k}_i|$,
and since $|{\bf k}_i|\approx Q\ll M_H$, we
find the momentum difference (note the opposite sign
with respect to $ \Delta  \omega_{{\bf k}}$)
%
%\begin{equation}
$ {\Delta |{\bf k}|}\
\equiv
{|{\bf k}_2|}
-{|{\bf k}_1|}\approx
(\lfrac{ \Delta m^2}{2M_H})-(\lfrac{ \Delta m^2}{2|{\bf k}|})\approx
-\lfrac{ \Delta m^2}{2Q}.
$
%\label{@DIFFK}\end{equation}
%
%This is of the same order as the large incoming pulsation frequency
%at the fixed
% average momentum
%$\bar {\bf k}=({\bf k}_1+{\bf k}_2)/2$.
%Still, both momenta
%are contained in the same wave packet
% of the outgoing antineutrino. This will be seen in Fig.~1.
}

\section{Experiments}

The best experimental results
are available
for the EC-processes
reported in Ref.~\cite{GSI2}, where  an
electron is captured from the K-shell
and converted into an electron-neutrino
which runs
 off to infinity.
On the average, the
decay
is exponential
with a
rate expected
from a standard-model calculation.
In addition, however,
the decay rate
shows modulations with a frequency $ \Delta  \omega $.
The experimental results
are \cite{rema}
\begin{eqnarray}\label{@EX1}
&&{}^{140}_{\phantom{1}59}{\rm Pr}^{58+}~\rightarrow\,
{}^{140}_{\phantom{1}58}{\rm Ce}^{58+}
:~\,\,\Delta  \omega\approx0.890(11)\,{\rm sec}^{-1}~~~(Q=3\,386\,{\rm keV}),\\
&&{}^{142}_{\phantom{1}61}{\rm Pm}^{60+}\rightarrow
{}^{142}_{\phantom{1}60}{\rm Nd}^{60+}
:~~\Delta  \omega\approx0.885(31)\,{\rm sec}^{-1}~~~(Q=4\,470\,{\rm keV}).
\label{@EX2}\end{eqnarray}
In both cases, the period of modulation is roughly $7$ sec, and scales with
$M$ (see Fig. 1).
The decay rate has the form
$ \lambda (t)= \lambda (0)(1+a\cos( \Delta  \omega t+ \Delta \phi)]$
with a modulation amplitude  of $a=0.18(3)$.
\begin{figure}
 \centering
	\includegraphics*[width=6cm,height=4.5cm]{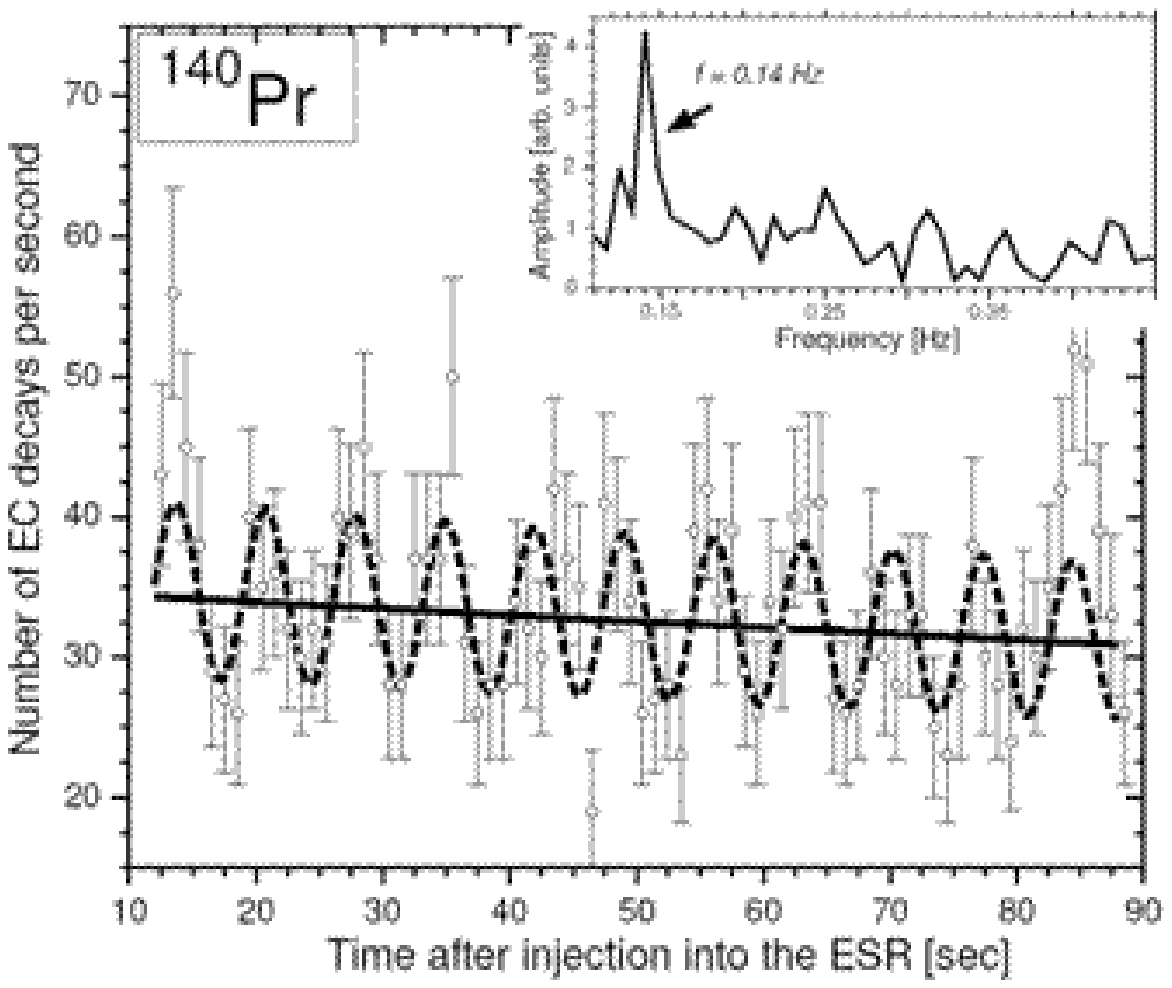}
	\includegraphics*[width=6cm,height=4.5cm]{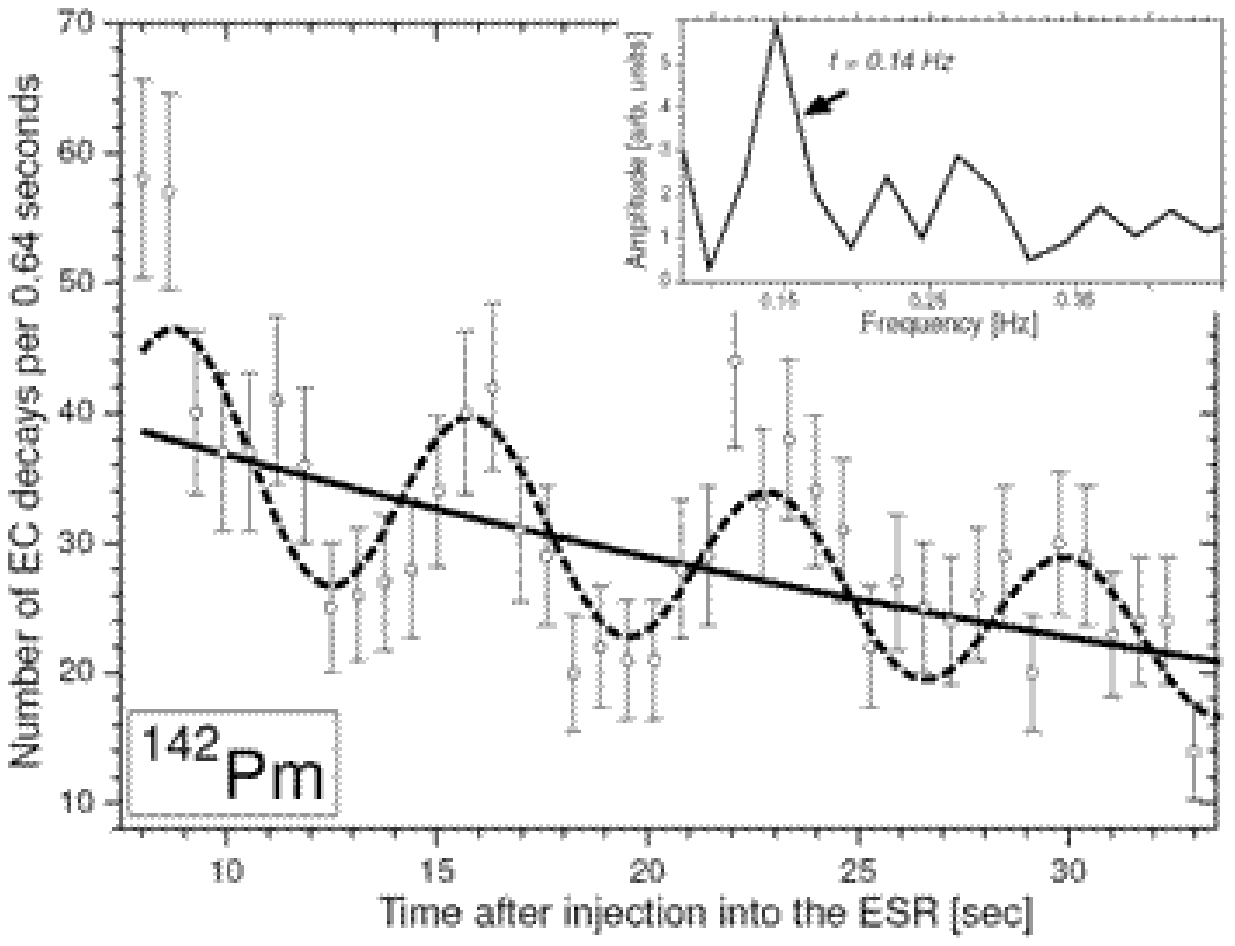}
\caption[]{Modulations of decay rate
for the processes
${}^{140}_{\phantom{1}59}{\rm Pr}^{58+}~\rightarrow\,
{}^{140}_{\phantom{1}58}{\rm Ce}^{58+}$ and
${}^{142}_{\phantom{1}61}{\rm Pm}^{60+}\rightarrow
{}^{142}_{\phantom{1}60}{\rm Nd}^{60+}$.
The period is in both cases roughly $7$ sec.
The inserts show the frequency analyses.
Plots are from
Ref.~\cite{GSI2}. The decay rate is modulated by a factor
$1+a\cos( \Delta  \omega  t+ \Delta \phi)$ with $a=0.18(3)$.
}
\label{figure}
\end{figure}

We expect theses modulations
 to
be associated with the frequency
$
\Delta  \omega$
of Eq.~(\ref{@EQUA}),
and thus to give information on
$ \Delta m^2$.
Inserting the experimental numbers
for $ \Delta  \omega $
into Eq.~(\ref{@EQUA}) and taking into account that the particles
in the storage ring run around with 71\% of the light velocity
with a Lorentz
factor
$ \gamma\approx1.43$,
we
 find
for both processes
 \cite{REM2}
\begin{equation}
 \Delta m^2\approx 22.5\times 10^{-5}\,{\rm eV}^2.
\label{@}\end{equation}
This is by a factor $\approx2.8$ larger than the result
$ \Delta m^2\approx7.58^{+0.3}_{-0.3}\times 10^{-5} {\rm eV}^2$
favored by the KamLAND experiment \cite{Kam,LIP}, but it lies close
to their less favored result \cite{KAMLex},
which the authors excluded
by 2.2$ \sigma$
in 2005, and now by 6$ \sigma $ \cite{Kam}
(see Fig. 2).
\begin{figure}
 \centering
	\includegraphics*[width=6cm]{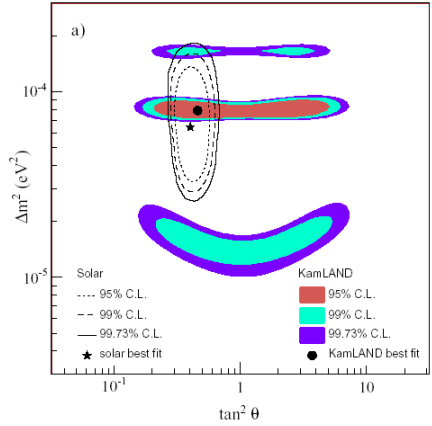}
\caption{The upper KamLAND regime of 2006 \cite{KAMLex} is compatible with
our result  $ \Delta m^2\approx22.5\times 10^{-5} {\rm eV}^2$.}
\label{figure}
\end{figure}

So far we do not yet understand
the origin of this discrepancy.
One explanation has been attempted in Ref.~\cite{IVN}
where
the authors
investigate the influence
of the strong Coulomb field
around the ion upon the process.

\section{Entangled Wavefunction}

For a theoretical explanation of the
modulations, we first simplify the situation
and
ignore all spins and the finite size of the ions.
Then the decay of
the initial ion $I$ into
the ion $I_H$ plus an electron-antineutrino $\bar  \nu _{e}$.
 can be described by an effective
interaction for this process is
\begin{eqnarray}
{\cal A}_{\rm int}=g
\int d^4x\,I_H^\dag(x)\nu_e(x)I(x)=g
\int d^4x\left[
\cos \theta\, I_H^\dag(x)\nu_1(x)I(x)
+\sin \theta\, I_H^\dag(x)\nu_2(x)I(x)\right],
\label{twot}\end{eqnarray}
where $I(x)$, $  \nu _e(x)$, and $I_H(x)$ are the field operators
of the involved particles.
In the CM frame, the initial ion is at rest, the final moves
nonrelativistically.
The outgoing wave is spherical.
The
role of the antineutrino
creation operators
in $  \nu _1$
and $  \nu _2$
is simply to create a coherent superposition
of two such waves
with the two different $k$- and $ \omega $-values
calculated above.
The combined outgoing wave function
will be
\begin{eqnarray} &&
\!\!\!\!\!\!\!\!\!\!\!\!\!\!\!
\!\!\!\!\!\!\!\!\!\!\!\!\!\!\!
\int \frac{d^3p}{(2\pi)^3}
\int \frac{d^3k}{(2\pi)^3}
\left[ ~
\cos \theta
e^{-iE_{\sbf p}t+i{\sbf p}{\sbf x}}|I_H({\bf p})\rangle
e^{-i \omega_{{\ssbf k},1}t+i{\sbf k}{\sbf x}}|\bar \nu _1({\bf k})\rangle
\right.\nonumber \\~~~~~~~~~~&&~~~~~~~~\left.+
\sin \theta
e^{-iE_{\sbf p}t+i{\sbf p}{\sbf x}}|I_H({\bf p})\rangle
e^{-i \omega_{{\ssbf k},2}t+i{\sbf k}{\sbf x}}|\bar \nu _2({\bf k})\rangle\right] .
\label{@11}\end{eqnarray}
The states
	    $|\bar \nu _1({\bf k})\rangle,
	    |\bar \nu _2({\bf k})\rangle $, in turn, can be
reexpressed in terms of the electron-
and muon-neutrino states as
\begin{eqnarray}
|\bar \nu _1({\bf k})\rangle=
\cos \theta|\bar \nu _e({\bf k})\rangle-
\sin \theta| \bar\nu _\mu({\bf k})\rangle,~~~~
|\bar \nu _2({\bf k})\rangle=
\sin \theta|\bar \nu _e({\bf k})\rangle+
\cos \theta| \bar\nu _\mu({\bf k})\rangle.
\label{@11}\end{eqnarray}
Thus we find for the transition to an
electron-neutrino of any momentum the effective action
\begin{eqnarray}
\int \frac{d^3k}{(2\pi)^3}
\langle \bar  \nu _e(-{\bf k})|{\cal A}_{\rm int}|0\rangle
= g   \int d^3x   \,
I_H(x)I(x)v_{\bar \nu _e}(x)
\label{@12}
\label{twotp}
\end{eqnarray}
with a spacetime-dependent potential
\begin{equation}
v_{\bar \nu _e}(x)=
\int \frac{d^3k}{(2\pi)^3}\left[
\cos^2 \theta e^{i \omega_{{\ssbf k},1}t+i{\sbf k}{\sbf x}}
+\sin^2 \theta e^{i \omega_{{\ssbf k},2}t+i{\sbf k}{\sbf x}}\right] .
\label{@13}\end{equation}
In Born approximation we find from this the scattering state
of the recoiling ion
$I_H$:
\begin{eqnarray}
\langle {\bf x}|\psi^{(+)};t\rangle^{\bar  \nu _e}\equiv
-\frac{g}{r}\left[ \cos^2 \theta\, e^{i(k_1r- \omega_1 t)}
+ \sin^2 \theta \, e^{i(k_2r- \omega_2 t )}\right].~~~
\label{@10}\end{eqnarray}
This
wave
carries  a radial current density
of ions $I_H$
\begin{eqnarray}
j^{\bar  \nu _e}_r=\frac{g^2}{M_Hr^2}\left[
\cos^4\theta\, k_1
+\sin^4\theta\, k_2 +
\sin^2\theta
\cos^2\theta \,(k_1+k_2)
\cos  ( \Delta k\,r-\Delta  \omega \,t)\right].
\label{@kbar}\end{eqnarray}
In order to find  the decay rate
we integrate this over a sphere
of radius $R$
surrounding the initial ion, choosing for
$R$ any size
$\ll 1/ \Delta k\approx 10^4\,$m.
For this surface we
find the
outgoing probability current density
\begin{equation}
\dot P=4\pi g^2\frac{\bar k}M \left[
1-\sfrac{1}{2}\sin^2(2\theta)
+\sfrac{1}{2}\sin^2(2\theta)
\cos  (\Delta  \omega \,t)\right],
\label{@16}\end{equation}
where we have approximated $k_1$ and $ k_2$ by their average $\bar k$.

This $\dot P$ can explain directly
the observed modulations of the decay rate
of the initial ions.
The is only one problem: the
amplitude
of modulations are predicted to be
$a=
\sfrac{1}{2}\sin^2(2\theta)/[1-
\sfrac{1}{2}\sin^2(2\theta)]\equiv 0.72$.
Experimentally, however, $a$ is
much smaller. It has the value $~0.18(3)$.
%due to the
%finite momentum width if the
%initial wave packet.

The discepancy is explained by
a missing contribution to the
deay rate.
So far we have only included
the contribution
of the effective action (\ref{twot}).
There is, however, also a second effective action which
is generated by the matrix elements
\begin{eqnarray}
\int \frac{d^3k}{(2\pi)^3}
\langle \bar  \nu _\mu(-{\bf k})|{\cal A}_{\rm int}|0\rangle
= g   \int d^3x   \,
I_H(x)I(x)v_{\bar \nu _\mu}(x)
\label{twotpp}
\label{@17}
\end{eqnarray}
where
\begin{equation}
v_{\bar \nu _\mu}(x)=
\int \frac{d^3k}{(2\pi)^3}
\sin\theta\cos\theta\left[ - e^{i \omega_{{\ssbf k},1}t+i{\sbf k}{\sbf x}}
+ e^{i \omega_{{\ssbf k},2}t+i{\sbf k}{\sbf x}}\right] .
\label{@18x}\end{equation}
 This can be derived directly
from Eqs.~(\ref{@11})
and (\ref{@12}).
Here the Born approximation yields
the scattering state
of the recoiling ion
$I_H$:
\begin{eqnarray}
\langle {\bf x}|\psi^{(+)};t\rangle^{\bar  \nu _\mu}\equiv
\frac{g}{r}\sin \theta \cos \theta\,
\left[ e^{i(k_1r- \omega_1 t)}
- e^{i(k_2r- \omega_2 t )}\right].~~~
\label{@10}\end{eqnarray}
Its radial
 current density
is now
\begin{eqnarray}
j^{\bar  \nu _\mu}_r=\frac{g^2}{M_Hr^2}
\sin^2\theta
\cos^2\theta(k_1+k_2)
\left[
1-
\cos  ( \Delta k\,r-\Delta  \omega \,t)\right].
\label{@}\end{eqnarray}
Now we have another problem: the modulations of this current cancel
the modulations of the current
(\ref{@16}).
We may suspect that this has to do with
the fact that there are three neutrinos which we must
take into consideration.

\section{Three-Neutrino Mixing}

Let us now include
all three
known neutrinos $ \nu _e, \nu _\mu, \nu _\tau $. Their
fields are denoted by
$ \nu _ \sigma $ with $ \sigma =e,\mu,\tau $.
These fields are combinations of three fields with definite mass
\begin{eqnarray}
  \nu _ \sigma =U_{ \sigma i} \nu _i,~~~~ \sigma =(e,\mu,\tau ),
\label{@}\end{eqnarray}
The mixing matrix $U_{ \sigma i}$ is called
Maki-Nakagawa-Sakata matrix, or short MNS-matrix, the neutrino analog of the
$3\times 3$ Cabibbo-Kobayashi-Maskawa matrix for the mixing of the quarks
$d,s,b$,
It is commonly assumed to be unitary, i.e., to satisfy the relation
\begin{equation}
\sum_ \sigma U^*_{ \sigma j}
U_{ \sigma l}= \delta _{jl}.
\label{@unit}\end{equation}
Its standard parametrization is
the following product of
 four simple
unitary matrices
\begin{eqnarray}
U=
\left(
\begin{array}{ccc}
1&0&0\\
0&c_{23}&s_{23}\\
0&-s_{23}&c_{23}
\end{array}
\right)
\left(
\begin{array}{ccc}
c_{13}&0&s_{13}e^{-i \delta }\\
0&1&0\\
-s_{13}e^{ i \delta }&0&c_{13}\\
\end{array}
\right)
\left(
\begin{array}{ccc}
c_{12}&s_{12}&0\\
-s_{12}&c_{12}&0\\
0&0&1
\end{array}
\right)
\left(
\begin{array}{ccc}
e^{i \alpha _1}&0&0\\
0&e^{i \alpha _2}&0\\
0&0&1
\end{array}
\right)
\label{@}\end{eqnarray}
where
$s_{ij}\equiv \sin \theta_{ij}$,
$c_{ij}\equiv \cos \theta_{ij}$.
For quarks, the unitarity
relation   (\ref{@unit})
is presently in the focus of
experimental and theoretical
studies in many research groups \cite{FORI}.
For leptons, the data have so far been insufficient
to test it.

Generalizing
(\ref{@12}),
(\ref{@13})
and
(\ref{@17}),
(\ref{@18x}),
we have from each flavor $ \sigma $ an effective potential
\begin{eqnarray}
\int \frac{d^3k}{(2\pi)^3}
\langle \bar  \nu _ \sigma (-{\bf k})|{\cal A}_{\rm int}|0\rangle
= g   \int d^3x   \,
I_H(x)I(x)v_{\bar \nu _ \sigma }(x)
\label{twotpx}\end{eqnarray}
with a potential
\begin{equation}
v_{\bar \nu _ \sigma }(x)=
\int \frac{d^3k}{(2\pi)^3}   \sum_{j=0}^3
U_{ej}
U^*_{ \sigma j}
 e^{i \omega_{{\ssbf k},j}t+i{\sbf k}{\sbf x}}
 .
\label{@18xy}\end{equation}
This produces
an outgoing wave function of the ion $I_H$
in the center-of-mass frame due to
the potential $v_{\nu _ \sigma }$ is
\begin{eqnarray}
\langle {\bf x}|\psi^{(+)};t\rangle^{\bar  \nu _ \sigma }\equiv
-\frac{g}{r}\sum_{j=1}^ 3
U_{e j }
U^*_{ \sigma  j }
e^{i(k_jr- \omega_j t)},
\label{@10}\end{eqnarray}
with an ion current density
\begin{eqnarray}
j^{\bar  \nu _ \sigma }_r=\frac{g^2}{M_Hr^2}
\sum_{j,l=1}^3
\sum_{ \sigma =1}^3
U_{ej}U^*_{ \sigma j}
U^*_{el}U_{ \sigma l} \,
k_j
e^{i[(k_j-k_l)r-( \omega _j- \omega _l)t]}
.\label{@27}\end{eqnarray}
If we sum over all flavors of the antineutrino
and use the unitarity relation (\ref{@unit}), we obtain
the total radial current density
\begin{eqnarray}
j_r\equiv \sum_ \sigma j^{\bar  \nu _ \sigma }_r=\frac{g^2}{M_Hr^2}
\sum_{j=1}^3
U_{ej}
U^*_{ej}
\,k_j
.\label{@}\end{eqnarray}
As previously for two flavors, the modulations
in (\ref{@27})
 disappear.
\comment{
If we ignore the mass difference
as in  Eq.~(\ref{@16}),
we can approximate
\begin{eqnarray}
\sum_ \sigma j^{\bar  \nu _ \sigma }_r\equiv \frac{g^2}{M_Hr^2}
\bar k\sum_{j=1}^3
U_{ej}
U^*_{ej}
.\label{@}\end{eqnarray}
}

However, the GSI experiments
did observe modulations
with an amplitude
$a\approx0.18(3)$.
Thus we must conclude
that the
unitarity
relation (\ref{@unit}) must be violated.
Since so far only the
lowest possible
modulation frequency $ \Delta  \omega = \omega _2- \omega _1$
 between the two lightest neutrinos
has been measured,
we may parametrize the right-hand side
of the unitarity violation
 by
\begin{equation}
\sum_ \sigma U^*_{ \sigma j}
U_{ \sigma l}=
u_0 \delta _{jl}+
u_{21}
%(\delta _{j,l+1}+\delta _{j,l-1})
(
\delta _{j,2} \delta _{l,1}
+\delta _{j,1} \delta _{l,2})
+\dots
,
\label{@unit2}\end{equation}
and find
\begin{eqnarray}
j_r%=\sum_ \sigma j^{\bar  \nu _ \sigma }_r
=\frac{g^2}{M_Hr^2}
\left\{ u_0S_0
 \delta _{ij}
+2u_{21} S_{21}
 \cos[ \Delta kr- \Delta  \omega t+  \Delta \phi]\right\}
,\label{@}\end{eqnarray}
where
\begin{equation}
S_0\equiv
\sum_{j=1}^3 U_{ej}U^*_{ej}k_j,~~~~
S_{21}\equiv  |U_{e\,2}U^*_{e1}|(k_{2}+k_1)/2,~~~~  \Delta \phi
\equiv \arg U_{e2}U^*_{e1}.
\label{@}\end{equation}
%
%The indices are understood to run
%cyclically through $1,2,3$, i.e., $k_{4}\equiv k_{1},k_{0}\equiv k_{3}$.
Assuming that the violation of unitarity
is small, the sums $S_0$ and $S_1$ are close to unity.
Then we deduce from tha
experimental result $a\approx0.18(3)$ that
\begin{equation}
\frac{u_{21}}{u_0}\approx 10\%\,.
\label{@}\end{equation}

A possible origin of this
unitarity violation
could be that there are more than three families of leptons in nature
and that universality of weak interaction is not valid for all
of them.
If the symmetry between quarks and leptons
of  the standard model
persists to higher energies,
we do not expect more than eight lepton families
to exist---
more than eight quark families
would ruin asymptotic freedom
and thus confinement.
Thus there is room for more than the three
quark and lepton families
observed so far. Indeed,
a fourth set of families is under intense discussion
\cite{HUNG}
in connection with the new accelerator
LHC at CERN.
 So far,  there are only weak bounds
on their masses from different sources \cite{Kam}:
\begin{eqnarray}
m_{t'}\ge 256 {\rm\,GeV},~~
m_{b'}\ge 128 {\rm\,GeV},  ~~
m_{\tau '}\ge 100.8 {\rm\,GeV},~~
m_{ \nu _{\tau '}}\ge 90.3 {\rm\,GeV}.~~
\label{@}\end{eqnarray}
If any of the heavier leptons
is coupled with a coupling constant
that does not fit into the CKM scheme,
unitarity will certainly be violated.
More data will be needed to decide precisely how.

\comment{
From a simple phenomenological mass formala
set up some time ago by Barut \cite{BAR},
which explained correctly the $\tau $-mass of $\approx1790$MeV,
we can estimate the
next lepton mass to be $\approx10.30$GeV, the one after this to be
37.179GeV.
A Phenomenological Mass Fomula for Leptons and Quarks
Progress of Theoretical Physics
Vol. 82 No. 6 (1989) pp. 1125-1132			Hiroshi Katsumori,
}

\section{Comments}

It is noteworthy that this analysis,
in which we extract the properties of the unobserved
antineutrino from the
behavior of the ion,
corresponds
precisely to the usual
entanglement analysis
of decay processes
such as
$\pi^0      \rightarrow  \gamma + \gamma $.
There
the measurement of the polarization of {\em one\/} photon
tells us immediately
the polarization properties of the other, unobserved photon.

A few comments are in place on several recent publications
\cite{CRIT1,CRIT2,CRIT3,CRIT4,CRIT5,CRIT6}
which deny the relation
between neutrino oscillations and the
nonexponential
decay
seen
in the
GSI experiment for various reasons.
In Ref. \cite{CRIT1}, the basic argument
is that
the antineutrino oscillations  set in {\em after\/} their emission,
so that they cannot
be observed in the GSI
experiment.
% which measures the
%decrease of the number of initial ions.
The present discussion shows that
although the forst part of
this argument is true, the conclusion
depends on the unitary assumption of the mixing matrix.
\comment{
although it fails to
give
the correct
explanation of the data \cite{BEATS}.
Indeed,
the GSI data  do {\em not\/}
care about
the
neutrino oscillations in their propagation
away from the decay center.
The neutrinos merely serve to give
the ion
a coherent
kick of two different momenta and energies.
What is measured
are
the resulting oscillations of the
ion wave function caused
by this kick.
\comment{
The existence of two closely lying nuclear states is suggested
whose beats are observed.
In \cite{CRIT2}, the argument is based on the use of Feynamn diagrams.
These, however,
cannot be applied in the small-time regime
where the
oscillations are observed
since they require
the limit $t\rightarrow \infty$
where
the oscillations are gone.
Only  diagrams in {\em spacetime\/}
involving
a {\em propagator matrix\/}
in $ \nu _1,\nu _2$ field space
with off-diagonal matrix elements
are applicable, and these reproduce the above-calculated oscillations.
}
\section{Discussion}
}

Finally we should point out that
similar
 oscillation phenomena
in the associate production
of particles together with an oscillating partner
have been proposed and controversially discussed
before by many authors
in
the production of muons
together with antineutrinos in the
decay $\pi^-\rightarrow \mu^-+\bar  \nu_ \mu$
 \cite{SRIVmu},
and
in the production
of $ \Lambda $ hyperons
together with
neutral Kaons \cite{SRIVLa},
In the latter case
the oscillation would come from a nonunitarity of the
quark mixing matrix,
which seems to be much smaller
than that of the
neutrino
 mixing matrix
reported here.

\comment{There is no space in this short communication
to explain all interesting
ideas
presented
the many related publications on this subject
\cite{CRIT3,CRIT4,CRIT5,CRIT6}.
}
	  ~\\

~\\[-3mm]Acknowledgment:
We are grateful to G\"unter Kaindl for
organizing regular scientific evening talks
in the Magnus-Haus Berlin of the German Physical Society.
This brought us together on March 11, 2008, in a lecture
by P.K. on the GSI experiments,
leading to the interpretation presented here.
We
also appreciate financial support by the
WE-Heraeus-Stiftung for the
'Nachsitzung' in the Remise of the Magnus-Haus.

\section{Appendix: Properties of Outgoing Wave}
The interaction is time-dependent
and we must adapt the scattering theory to this situation.
Recall briefly the
theory for a time-independent
interaction,  where
the scattering amplitude
is
obtained from
the standard limiting formula
\begin{eqnarray} \label{7.6.178}
  \langle {\bf p}' | \hat S|{\bf p}\rangle
\!&=&\!\lim_{t\to {\infty} } \langle {\bf p}' \vert {\bf p}^{(+)}(t)\rangle
\!=\!\lim_{t\to {\infty} } \langle {\bf p}' \vert {\hat U}_I(t,0)\vert {\bf p}^{(+)}\rangle
\! =\!
 \lim_{t \to {\infty} } \langle {\bf p}'
	  \vert e^{i{\hat H}_0 t} e^{-i({\hat H}_0+
\hat V) t} \vert {\bf p}^{(+)}
\rangle
\nonumber \\        &=&  \lim_{t \to {\infty} }
 e^{i( E_{{\ssbf p}'}-E_{\ssbf p})
	      t} \langle {\bf p}' \vert {\bf p}^{(+)} \rangle.
\end{eqnarray}
Here  $
\vert {\bf p}'\rangle
$ denotes an eigenstate
 of the
free
Hamiltonian
$\hat H_0$
with momentum ${\bf p}'$ and energy $E_{{\sbf p}'}={\bf p}'^2/2M$,
and
$\vert {\bf p}^{(+)}\rangle$
%\vert {\bf p}^{(+)}\rangle e^{-i E_{\sbf p}t}
 is an eigenstate
 of the interacting
Hamiltonian
$\hat H_0+\hat V$
with momentum ${\bf p}$ and energy $E_{\sbf p}$. It
solves the
Lippmann-Schwinger
equation:
\begin{equation}
  \vert {\bf p}^{(+)} \rangle  = \vert {\bf p} \rangle  + \frac{1}
  {E_{\sbf p}-{\hat H}_0 +i\eta } {\hat V} \vert {\bf p}^{(+)} \rangle ,
\label{@LS0}\end{equation}
which is verified by multiplying both sides by $E-\hat H_0$ from the left.
Inserting this into
(\ref{7.6.178})
leads to
\begin{equation} \label{7.6.179}
  \langle {\bf p}' \vert \hat S \vert {\bf p} \rangle  =
\langle {\bf p}' \vert {\bf p}\rangle
 +
      \lim_{t \to {\infty} }
  \frac{
e^{i( E_{{\ssbf p}'}-
E_{\ssbf p}) t}
}{E_{\sbf p} - E_{{\sbf p}'} + i\eta }
	     \langle {\bf p}' \vert {\hat V} \vert {\bf p}^{(+)}
   \rangle ,
\end{equation}
 where $ \eta >0$ is an
infinitesimally number.
The second term
%in (\ref{7.6.179})
contains
the $T$-matrix $ T_{{\sbf p}',{\sbf p}}\equiv \langle {\bf p}' \vert {\hat V} \vert {\bf p}^{(+)}\rangle
$
which describes
true scattering.
In the absence of neutrino oscillations, $|{\bf p}\rangle $ is simply
the initial ion at rest,
and $ \langle {\bf p}' \vert$ the state with the ion $I_H$
with momentum ${\bf p}+{\bf k}$ and the antineutrino $\bar  \nu _e$
with momentum ${-\bf k}$.
The
 limit $t \rightarrow {\infty} $
in the prefactor
can simply be taken after rewriting
it as
%
%\begin{eqnarray}
$-i\int_{-\infty}^{t } dt    e^{i( E_{{\ssbf p}'}-E_{\ssbf p}-i \eta) t},
$ %\label{@}\end{eqnarray}
which obviously tends to $-2\pi i \delta (E_{{\ssbf p}'}-E_{\ssbf p})$ in the limit
$t\rightarrow \infty$.
The $\delta $-function ensures the conservation of energy
in the process.
This is, of course, the standard derivation of {\em Fermi's Golden Rule\/}
which we repeated here
to clarify
that it is applicable only
to processes in which the final state is an eigenstate of the
free-particle Hamiltonian operator $\hat H_0$.

Another way of deriving this result
is based on the
spatial wave function associated with the
state $|{\bf p}^{(+)}\rangle$. One multiplies Eq.~(\ref{@LS0})
by the state $\langle {\bf x}|$
from the left and
obtains the wave funcion
\begin{equation}
\langle {\bf x}  \vert {\bf p}^{(+)} \rangle  =
\langle {\bf x} \vert {\bf p} \rangle  +\int d^3x'G(E_{\sbf p};{\bf x},{\bf x}')
{\hat V}({\bf x}')\langle {\bf x}'| {\bf p}^{(+)} \rangle ,
\label{@LS1}\end{equation}
where
\begin{eqnarray}
G(E;{\bf x},{\bf x}')
\equiv
\langle {\bf x} |\frac{1}{
{E-{\hat H}_0 +i\eta }}
|{\bf x}'\rangle= \int
\frac{d^3p'}{(2\pi)^3}
\frac{e^{i{\sbf p}'({\sbf x}-{\sbf x}')}}
 {E-{\bf p}'{}^2/2M +i\eta }
\approx
-2M\,\frac{e^{i{ p}'_{E}r}}
{4\pi r}
e^{-i{ p}'_{E}\hat {\sbf x}{\sbf x}'}
,\label{@}\end{eqnarray}
with
$
r\equiv |{\bf x}-{\bf x}'|$,
\,$\hat {\bf x}\equiv
 {\bf x}/r
$
and
$
{ p}'_{E}= \sqrt{2M E}
$.

In Born approximation, one inserts
on the right-hand side of
 Eq.~(\ref{@LS1}) a plane wave
$
\langle {\bf x}'| {\bf p}^{(+)} \rangle
\approx\langle {\bf x}'| {\bf p} \rangle=e^{i
{\sbf x}'
{\sbf p}
}$,
and  Eq.~(\ref{@LS1})
becomes
\begin{equation}
\langle {\bf x}  \vert {\bf p}^{(+)} \rangle  =
\langle {\bf x} \vert {\bf p} \rangle
-2M\,\frac{e^{ip'_{E_{\ssbf p}}r}}{4\pi r}\int d^3x'
e^{-i({{\sbf p}}'-{\sbf p}){\sbf x}'}V({\bf x}'),
\label{@LS2}\end{equation}
where $  {\bf p}'$ is short for
the momentum of the outgoing particle
of energy $E_{\bf p}$  in the direction
of ${\bf x}$:
${\bf p}'\equiv
p'_{E_{\sbf p}}
\hat {\bf x}$.
Thus, in Born approximation,
the amplitude
for the final particle to  emerge with momentum
${\bf p}'$ is proportional the Fourier
transform of the potential at the momentum transfer
$ \Delta {\bf p}\equiv
{\bf p}'-
{\bf p}$.
If the potential
is a plane wave
of momentum $-{\bf k}$,
i.e.,
if
\begin{equation}
V({\bf x})=\frac{g}{(2\pi)^3}e^{i{\bf k}{\bf x}}
\label{@}\end{equation}
then the
final state has the wave function
\begin{equation}
\langle {\bf x}  \vert {\bf p}^{(+)} \rangle  =
\langle {\bf x} \vert {\bf p} \rangle
-2Mg\,
\frac{e^{ip'r}}{4\pi r} \delta ^{(3)}({\bf p}'-{\bf p}+{\bf k}).
\label{@LS2'}\end{equation}

Let us adapt this formalism
 to the oscillating situation.
According to Eqs.~(\ref{twotp}), (\ref{@13}),
the emission of an antineutrino of mass $m_1$
and momentum $-{\bf k}$
is described by
the time-dependent interaction potential
\begin{eqnarray}
v_{\bar nu_e}({\bf x},t)=\frac{e^{i\sbf k
\sbf x}}{(2\pi)^3}
e^{i
\omega_{\ssbf k,1}
 t
} ,~~~~~~
\omega_{\ssbf k,1}
= \sqrt{{\bf k}^2+m_1^2},
\label{@VV}\end{eqnarray}
where we have dropped the factor
$\cos^2 \theta$ accompanying
the
coupling  $g$, for brevity.
As before, the incoming ion $I$ has the momentum
${\bf p}$.
Its energy is ${\bf p}^2/2M$.
The outgoing ion $I_H$
has the momentum ${\bf p}'$ and an energy
$E_{{\sbf p}'}=  M_H-M+
{\bf p}'^2/2M_H$.

Let us first adapt
the
Lippmann-Schwinger approach.
We introduce
the time-dependent interacting state
$|{\bf p}^{(+)}(t)\rangle$, which
is an eigenstate of the full Hamiltonian, and satifies the time-dependent
Schr\"odinger equation
\begin{equation}
i\partial _t |{\bf p}^{(+)}(t)\rangle=
[\hat H_0+\hat v({\bf x},t)]|{\bf p}^{(+)}(t)\rangle,~~~~~\hat H_0=\hat{\bf p}^2/2M.
\label{@}\end{equation}
The formal solution
of this is
\begin{equation}
  \vert {\bf p}^{(+)}(t) \rangle
 \equiv
\hat U(t)
 \vert {\bf p} \rangle ,~~~~~\hat U(t)\equiv \hat Te^{-i\int _0^t dt'
[\hat H_0+\hat V({\bf x},t)]},
\label{@}\end{equation}
An implicit exression for this state
can be written,
by analogy with
(\ref{@LS0}), as
\begin{equation}
  \vert {\bf p}^{(+)}(t) \rangle  =\vert {\bf p} \rangle
e^{-iE^0_{\ssbf p}t}
 + \frac{1}
  {i\partial  _t-{\hat H}_0 +i\eta } {\hat V} ({\bf x},t)
 \vert {\bf p}^{(+)} (t)\rangle , ~~~~E^0_{\sbf p}={\bf p}^2/2M.
\label{@Lest}\end{equation}
This can again
be verified by multiplication from the left
with
$i\partial _t-\hat H_0$.
Multiplying
(\ref{@Lest})
by
$
e^{iE^0_{\ssbf p}t}
$, we obtain
\begin{equation}
e^{iE^0_{\ssbf p}t}
  \vert {\bf p}^{(+)}(t) \rangle  =\vert {\bf p} \rangle  + \frac{1}
  {i\partial _t+E^0_{\sbf p}-{\hat H}_0 +i\eta }e^{iE^0_{\ssbf p}t} {\hat V}
({\bf x},t) e^{-iE_{\ssbf p}t}\vert {\bf p}^{(+)} (0)\rangle ,
\label{@48}\end{equation}
To lowest approximation, we replace
$\vert {\bf p}^{(+)} (0)\rangle$ by
$\vert {\bf p}\rangle$ and insert (\ref{@VV})
to find
\begin{equation}
e^{iE^0_{\ssbf p}t}
  \vert {\bf p}^{(+)}(t) \rangle  =\vert {\bf p} \rangle  +g\,
\int_{-\infty}^\infty dt'\,\hat G(t,t')
%\int \frac{d^3k}{(2\pi)^3}
e^{i{\sbf k}\hat{\sbf x}}
e^{i \omega _{{\ssbf k},1}(t'-t_0)}
\vert {\bf p}\rangle e^{-i(E_{\ssbf p}-E^0_{\ssbf p})t'},
\label{@37}\end{equation}
where
$\hat G(t,t')$ is the Fourier representation of the
operator $(
{i\partial _t+E^0_{\sbf p}-{\hat H}_0 +i\eta })^{-1}$:
\begin{equation}
\hat G(t,t')\equiv
\int _{-\infty}^\infty\frac{dE}{2\pi}\frac{e^{-iE(t-t')} }
  {E+E^0_{\ssbf p}-{\hat H}_0 +i\eta }
\label{@}\end{equation}
Performing the integral over $t'$ in (\ref{@37})
yields
\begin{equation}
e^{iE^0_{\ssbf p}t}
  \vert {\bf p}^{(+)}(t) \rangle  =\vert {\bf p} \rangle  +g\,
%\int \frac{d^3k}{(2\pi)^3}
\frac{e^{-i(E_{\ssbf p}
-
E^0_{\ssbf p}- \omega _{{\ssbf k},1}
)t}}
{E_{\sbf p}-{\hat H}_0 -\omega _{{\ssbf k},1}+i\eta}
e^{i{\sbf k}\hat{\sbf x}}
\vert {\bf p}\rangle
e^{-i \omega _{{\ssbf k},1}t_0}
.
\label{@37a}\end{equation}
We  multiply this equation from the left
and insert in front of the right-hand state $|{\bf p}\rangle$
a completeness relation $\int d^3x|
{\bf x}'\rangle \langle
{\bf x}' |=1$.
Approximating the matrix elements
$\langle {\bf x} |(
{E_{\sbf p}-{\hat H}_0
 +i\eta })^{-1}|{\bf x}'\rangle$
as
usual in the large-${\bf x}$ regime
by
\begin{eqnarray}
\langle {\bf x} |\frac{1}{
{E_{\sbf p}-{\hat H}_0-\omega _{{\sbf k},1} +i\eta }}
|{\bf x}'\rangle= \int
\frac{d^3p'}{(2\pi)^3}
\frac{e^{i{\sbf p}'({\sbf x}-{\sbf x}')}}
 {E_{\sbf p}-{\bf p}'{}^2/2M-\omega _{{\sbf k},1} +i\eta }
\approx -2M\frac{e^{i
{ p}'_kr}}{4\pi r}
e^{-i{ p}'_k\hat{\sbf x}{\sbf x}'}
,\label{@}\end{eqnarray}
where $p'_k$ is the momentum
of the ion $I_H$ which conserves the energy, i.e.,
$p'_k{}^2/2M=E_{\sbf p}- \omega _{{\sbf k},1}$.
With this we
obtain
\begin{equation}
\langle {\bf x} |e^{iE^0_{\ssbf p}t}
  \vert {\bf p}^{(+)}(t) \rangle  =\langle {\bf x}\vert {\bf p} \rangle - 2M
\frac{g}{(2\pi)^3}\frac{e^{i
{ p}'_kr}}{4\pi r}
%\int \frac{d^3k}{(2\pi)^3}
\int {d^3x'}
e^{-i{p}'_k\hat{\sbf x}{\sbf x}'}
e^{i({\sbf p}+{\sbf k}){\sbf x}'}
%e^{-i(E_{\ssbf p} -E^0_{\ssbf p}- \omega _{{\ssbf k},1})t}
e^{-i \omega _{{\ssbf k},1}t_0}
.
\label{@37c}\end{equation}
We now perform
 the  integral over ${\bf x}'$.
For this we assume the inital state to have zero momentum,
${\bf p}+{\bf k}=0$.
The integral
over ${\bf x}'$  forces
the momentum
of the outcoming
ion  $I_H$ to be equal to
${\bf k}$.
The integral over ${\bf x}'$
creates
a $ \delta $-function
$(2\pi)^3 \delta ^{(3)}(p'_k\hat{\bf x}-
{\bf p}
-{\bf k}
)$,
so that
we obtain
\begin{equation}
\langle {\bf x} |e^{iE^0_{\ssbf p}t}
  \vert {\bf p}^{(+)}(t) \rangle  =\langle {\bf x}\vert {\bf p} \rangle -
2M\frac{g}{4\pi r}
e^{-i{p}'_k\hat{\sbf x}{\sbf x}'}
\delta ^{(3)}({\bf p}'-{\bf p}+{\bf k})
%e^{-i(E_{\ssbf p}-E^0_{\ssbf p}- \omega _{{\ssbf k},1})t}
e^{-i \omega _{1}t_0}
.
\label{@37d}\end{equation}
Note that
since
energy and momentum
are balanced, then
$\omega _{{\ssbf k},1}
=
\omega _{1}$ of Eq.~(\ref{@OmEn}).

Consider now the case of two oscillating mass states
and let us
study the temporal behavior of
the emerging energy distribution.
The experiment does not
explore the limit of very large times
but measures the $t$-dependence
starting from small $t$ after the ion enters the storage ring.
Instead of the limiting energy conservation
$ \delta $-function  $-2\pi i \delta(
E_{{\sbf p}'}-E_{\sbf p}) $
in (\ref{7.6.179}),
it observes an approximation to it
valied for short times. To find it
we insert,
instead of (\ref{@VV}),
 the mixed potential
(\ref{@13}) into Eq.~(\ref{@48}), so that the time-dependent factor in
the resulting
equation of type
(\ref{@37})
has the form
\begin{eqnarray}
-i\int_{0}^{t } dt \,  \left[
\cos ^2\theta\,  e^{i( E_{1,{\ssbf k}'}-E_{\ssbf k} ) t}
+\sin ^2\theta\, e^{i( E_{2,{\ssbf k}'}-E_{\ssbf k} ) t}\right] ,
\label{@EQ}\end{eqnarray}
 where $E_{i,{\sbf k}'}\equiv  \sqrt{{\bf k}'^2+M_H^2} +
\omega _{{\sbf k}',i}$
and
 $E_{{\sbf k}}=M_H+Q$.
Since $m_i^2\ll Q\ll M_H$, we can approximate
$  E_{i,{\sbf k}'}-E_{\sbf k}\approx  \omega' - \omega _i$
where $ \omega _i\approx Q+m_i^2/2M_H $.
Let us write $ \omega _{1,2}\equiv
 \bar  \omega \mp
\sfrac{1}{2}\bar  \Delta m^2/2M_H=
 \bar  \omega \mp
\sfrac{1}{2}\bar  \Delta  \omega $.
Then (\ref{@EQ})
becomes, with the abbreviations
 $C\equiv \cos ^2\theta$
and $S\equiv \sin^2 \theta$,
\begin{eqnarray}
&&\!\!\!\!\!\!\!\!\!\!\!\!\!\!\!\!\!\!\!\!\!\!\!\!\!\!\!\!-\left( \!
C\frac{
e^{i(  \omega' - \omega _1 ) t}-1}
{ \omega' - \omega _1 }
\!+S\frac{
e^{i(  \omega' - \omega _2 ) t}-1}
{ \omega' - \omega _2}\right)\nonumber \\&=&
-i\left\{ \!
C
e^{i(  \omega' - \omega _1 ) t/2}
\frac{ \sin  [(  \omega'\! - \omega _1 ) t/2]}
{( \omega' \!- \omega _1)/2}
\!+S
e^{i(  \omega' - \omega _2 ) t/2}\frac{
[\sin(  \omega'\! - \omega _2 ) t/2]}
{ (\omega'\! - \omega _2)/2 }\right\}
.\label{@EQ1}\end{eqnarray}
The absolute square of (\ref{@EQ1})  mutiplied by some factor
 $|
 T|^2$
determines probability $P(t)$
to find the initial ions in the ring
at the  time $t$.
The integral over the final momenta
is dominated by the immediate neighborhood of the poles
at $ \omega _{1,2}$ where $ |{\bf k}|\equiv Q$. There we
 may
ignore the
${\bf k}$-dependence of
$| T|^2$,  approximating
it by a constant,
 and obtain the
probability for the ion $I_H$ to emerge with an energy
$M_H+Q -\omega '$
in the center-of-mass frame
\comment{
\begin{eqnarray} \!\!\!\!
P^{ \omega '}(t)\approx \left\{
C^2
\frac{ \sin^2  [(  \omega' \!- \omega _1 ) t/2]}
{( \omega'\! - \omega _1)^2/4}
\!+\!S^2
\frac{ \sin^2 [ (  \omega'\! - \omega _2 ) t/2]}
{( \omega'\! - \omega _2)^2/4}
\!+\!  2CS \cos(\bar \Delta  \omega t/2)
\left(\frac{ \sin  [(  \omega' \!- \omega _1 ) t/2]}
{( \omega'\! - \omega _1)/2}
 \frac{\sin  [(  \omega'\! - \omega _2 ) t/2]}
{( \omega'\! - \omega _2)/2}\right)\right\} |T|^2.
\label{@3}
\end{eqnarray}
}
\begin{eqnarray} \!\!\!\!
P^{ \omega '}(t)\approx \left\{
C^2 s_1^2( \omega ')
\!+\!S^2s_2^2( \omega ')
\!+\!  2CS \cos(\bar \Delta  \omega t/2)
s_1( \omega ')
s_2( \omega ')
\right\} |T|^2,
\label{@3}
\end{eqnarray}
where
$s_i( \omega ')\equiv \sin [( \omega '- \omega _i)t/2]/[( \omega '- \omega _i)t/2]$.
For large $t$, the limiting relation
$\sin^2
a t/a^2\rightarrow t\pi  \delta (a)$ allows us
approximate
$s^2_i( \omega ')\approx 2\pi t  \delta ( \omega '- \omega _i)$,
and thus the first two terms in (\ref{@3})
by
\begin{equation}P^{ \omega '}_{12}(t)\approx2\pi t [
C^2 \delta ( \omega '- \omega _1)
+S^2 \delta ( \omega '- \omega _2)] |T|^2.
\label{@}\end{equation}
If this is integrated
over
$\int d^3k'/(2\pi)^3\approx Q^2\int d \omega '/2\pi^2$,
the probabilities
of   $\nu _1$- and
$ \nu _2$-decays simply add, thereby
yielding the ordinary $ \beta $-decay rate
$I\rightarrow I_H+\bar \nu _e$ without mixing.
%%%%%
Consider
now the third term in (\ref{@3}).
Here the integral over all $ \omega '$ yields
\begin{eqnarray}
\int d \omega ' P_{3}^{ \omega '}(t)\approx 2CS
\cos (
\bar  \Delta  \omega t/2) \,
2\pi  \frac{\sin(
 \bar \Delta  \omega t/2)}{\bar \Delta  \omega /2}
|T|^2.
\label{@}\end{eqnarray}
Thus we obtain for the total decay rate as a function of time
\begin{equation}
\dot P(t) =
\int d \omega ' \left[
\dot P_{12}^{ \omega '}(t)+
\dot P_{3}^{ \omega '}(t)  \right]
\approx2\pi\left[ 1+2CS \cos ( \bar\Delta  \omega t)\right] |T|^2.
\label{@}\end{equation}
%
%
%
%
%%%%%%%%%%%%
%The size of the oscillations decreases with time.
%This decrease is determined by the
%momentum dependence of the $T$-matrix.
%%%%%%%%%%%%%
It would be interesting to observe experimentally
the
predicted distribution
(\ref{@3}) of antineutrino energies
$ \omega '$
by measuring the recoil momenta ${\bf k}$
of the final ions $I_H$.
The distribution consists of two peaks associate with the
emission of the
antineutrinos
$\bar\nu_1$ and $\bar\nu_2$. Centered between them
lies the oscillating
distribution
proportional
to
$
s_1( \omega ')
s_2( \omega ')$
shown in
Fig. 1.
\comment
{
%
% Creating boxdia.tex for boxdia.eps

\unitlength1mm
\def\fsz{\footnotesize}
\def\ssz{\scriptsize}
\def\tsz{\tiny}
\def\dst{\displaystyle}
\def\pu#1#2{\put(#1,#2){\emmoveto}}
\def\pd#1#2{\put(#1,#2){\emlineto}}
%%%%%%%%%%%%%%%%%%%%
\begin{figure}[tbhp]
\begin{picture}(105.64,29.645)
\def\dst{\displaystyle}
\def\fsz{\footnotesize}
\put(0,0){\includegraphics[height=3cm]{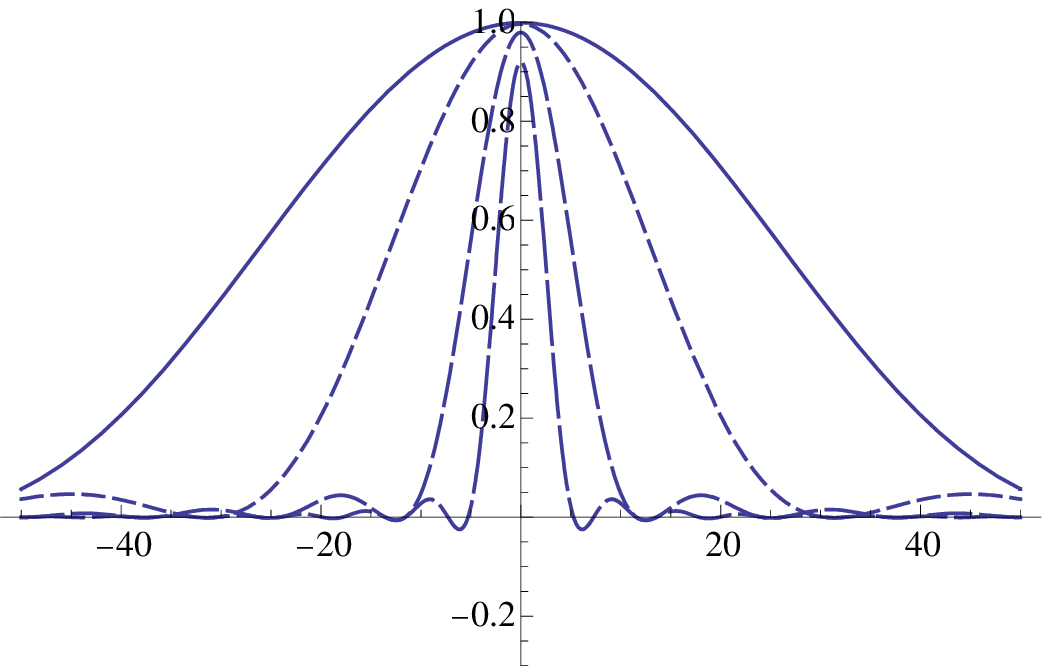}}
%\put(0,0){\input c:/emtex/texdraw/gratingf}
\put(32,25){\tsz $ t=0.1/ \Delta  \omega$}
\put(30,20){\tsz $ t=0.2/ \Delta  \omega$}\put(27,16){\tsz $ \Delta  \omega t=0.5/ \Delta  \omega$}
\put(25.2,12){\tsz $  t=1.0/ \Delta  \omega$}
\put(49,6.5){\tsz $  \omega '/\Delta  \omega $}
\end{picture}
\caption[]{
Distribution of the
temporally oscillating part of the antineutrino center-of-mass energies
$ \omega '$ in the decay $I\rightarrow I_H+\bar  \nu_e $
for different times $t$
[the function $
s_1( \omega ')
s_2( \omega ')$
in Eq.~(\ref{@3})].
}
\label{@}\end{figure}
}
\begin{figure}
 \centering
	\includegraphics*[width=6cm,height=4.5cm]{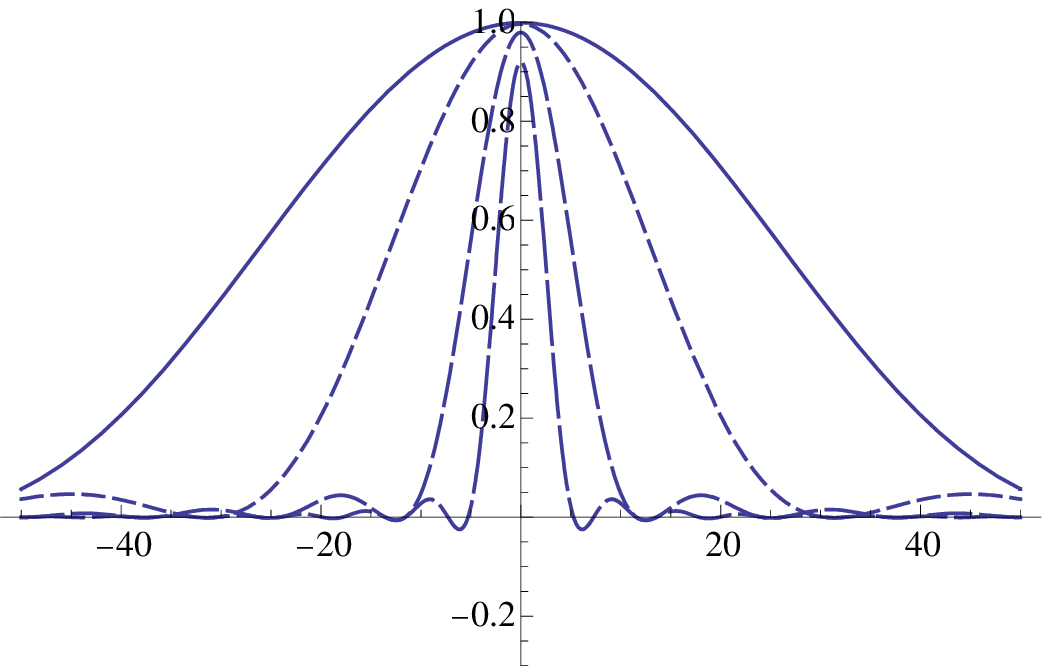}
\caption[]{Distribution of the
temporally oscillating part of the antineutrino center-of-mass energies
$ \omega '$ in the decay $I\rightarrow I_H+\bar  \nu_e $
for different times $t$
[the function $
s_1( \omega ')
s_2( \omega ')$
in Eq.~(\ref{@3})].
}
\label{figure}
\end{figure}

Note that the usual
Feynman diagrams in momentum space
cannot be used to
describe
the
observed oscillations
as done in Ref.~\cite{CRIT2}, since they imply
taking the limit $t\rightarrow \infty$
in which
the oscillations disappear.
Only  diagrams in {\em spacetime\/}
involving
a {\em propagator matrix\/}
in $ \nu _1,\nu _2$ field space
with off-diagonal matrix elements
are applicable, and these reproduce the above-calculated oscillations.


\begin{thebibliography}{99}
\bibitem{REM2}

The isotope masses of the final ions are
$M\approx130.319$\,GeV and
 $M\approx132.186$\,GeV, for
${}^{140}_{\phantom{1}58}{\rm Ce}$
and ${}^{142}_{\phantom{1}60}{\rm Nd}$, respectively.
Also recall that
a frequency
 $1/{\rm sec}$ corresponds to $\approx
 6.6\times 10^{-16}$\,eV.
\bibitem{REMA}
This cancellation was not noted in either of the previous
analyses
of Ref.~\cite{LIP}.
\end{thebibliography}
\end{document}